\documentclass[10pt]{article}
\usepackage{latexsym, epsfig}
\usepackage{amsmath}
\usepackage{amsfonts}
\usepackage{url}

\newtheorem{theorem}{Theorem}
\newtheorem{lemma}{Lemma}

\newcommand{\qed}{\rule{7pt}{7pt}}
\newenvironment{proof}{\noindent{\bf Proof}\hspace*{1ex}}{\qed\bigskip}

\def\calH{{\cal H}}

\def\calL{{\cal L}}
\def\calM{{\cal L}}

\def\yes{{\rm yes}}
\def\no{{\rm no}}
\newcommand{\CC}{\mathbb{C}}
\newcommand{\prob}[1]{\mathbb{P}{#1}}
\newcommand{\bea}{\begin{eqnarray} }
\newcommand{\ex}[1]{\mathbb{E}{#1}}

\newcommand{\eea}{\end{eqnarray} }
\newcommand{\tr}{\mathop{\mathrm{Tr}}}
\newcommand{\nn}{\nonumber}
\newcommand{\lrg}{{\mbox{\rm LARGE}}}
\newcommand{\sml}{{\mbox{\rm SMALL}}}
\newcommand{\acc}[1]{\mathrm{acc}(M,#1)}
\newcommand{\paths}[1]{\mathrm{path}(M,#1)}

\newcommand{\QMA}{\ensuremath{\mathrm{QMA}}}
\newcommand{\MA}{\ensuremath{\mathrm{MA}}}
\newcommand{\AM}{\ensuremath{\mathrm{AM}}}
\newcommand{\NP}{\ensuremath{\mathrm{NP}}}
\newcommand{\BPP}{\ensuremath{\mathrm{BPP}}}
\newcommand{\BPPpath}{\ensuremath{\mathrm{BPP_{path}}}}
\newcommand{\POSTBPP}{\ensuremath{\mathrm{PostBPP}}}

\renewcommand{\P}{\ensuremath{\mathrm{P}}}

\newcommand{\ba}{\begin{array}}
\newcommand{\ea}{\end{array}}

\def\be{\begin{equation}}
\def\ee{\end{equation}}
\def\bea{\begin{eqnarray}}
\def\eea{\end{eqnarray}}
\newcommand{\ra}{\rangle}
\newcommand{\la}{\langle}
\newcommand{\ket}[1]{| #1\rangle}
\newcommand{\bra}[1]{\langle #1|}

\newcommand{\ketbra}[1]{|#1\rangle\langle #1|}

\newcommand{\slno}{\Rightarrow}

\newtheorem{defi}[theorem]{Definition}

\newcommand{\ignore}[1]{}

\newcommand{\htar}{H_{\rm target}}
\newcommand{\helse}{H_{\rm else}}

\begin{document}

\title{The Complexity of Stoquastic Local Hamiltonian Problems}

\author{Sergey Bravyi \thanks{IBM Watson Research Center, Yorktown Heights, NY, USA 10598.
\texttt{sbravyi@us.ibm.com}} \and David P. DiVincenzo \thanks{IBM
Watson Research Center, Yorktown Heights, NY, USA 10598.
\texttt{divince@watson.ibm.com}} \and Roberto Oliveira \thanks{IBM
Watson Research Center, Yorktown Heights, NY, USA 10598.
\texttt{rob.oliv@gmail.com}} \and Barbara M. Terhal \thanks{IBM
Watson Research Center, Yorktown Heights, NY, USA 10598.
\texttt{bterhal@gmail.com}}}

\maketitle

\begin{abstract}
We study the complexity of the Local Hamiltonian Problem (denoted as
LH-MIN) in the special case when a Hamiltonian obeys the condition
that all off-diagonal matrix elements in the standard basis are real
and non-positive.  We will call such Hamiltonians, which are common
in the natural world, {\em stoquastic}.  An equivalent
characterization of stoquastic Hamiltonians is that they have an
entry-wise non-negative Gibbs density matrix for any temperature. We
prove that LH-MIN for stoquastic Hamiltonians belongs to the
complexity class \AM{}
--- a probabilistic version of \NP{} with two rounds of
communication between the prover and the verifier. We also show that
$2$-local stoquastic LH-MIN is hard for the class \MA. With the
additional promise of having a polynomial spectral gap, we show that
stoquastic LH-MIN belongs to the class \POSTBPP=\BPPpath
--- a generalization of \BPP{} in which a post-selective readout is
allowed. This last result also shows that any problem solved by
adiabatic quantum computation using stoquastic Hamiltonians is in
${\rm PostBPP}$.
\end{abstract}

\section{Introduction}
\label{sec:intro}

For the last few years significant progress has been made in
understanding the computational complexity of spin Hamiltonian
problems. This area of research is of great importance for physics,
since most strongly interacting quantum many-body systems can not be
fully analyzed by analytical methods; thus, we can only hope to
understand their properties from numerical simulations.  A system is
efficiently simulatable if the computational resources one needs for
simulation grow only polynomially with the number of spins in the
system. For example, one-dimensional spin chains with a small amount
of entanglement can be simulated by the DMRG method and its recent
generalizations to matrix product
states~\cite{White:DMRG,Rommer:DMRG,VPC:DMRG}. It has been proposed
that systems of interacting bosons, like those described by the
bosonic Hubbard model, can be simulated using the Green's function
Monte-Carlo technique, see~\cite{Trivedi:GFMC,Buonaura:GFMC}. It is
believed that a quantum computer will offer more possibilities to
simulate quantum systems. Understanding the computational complexity
of spin Hamiltonian problems might help to identify classes of
Hamiltonians for which efficient classical or quantum simulation
algorithms could be developed.

We shall consider the Local Hamiltonian Problem defined
in~\cite{KSV:computation,KKR:hamsiam}. A $k$-local $n$-qubit
Hamiltonian is a Hermitian operator $H$ acting on $(\CC^2)^{\otimes
n}$ that can be expressed as a sum of $k$-qubit interactions:
$H=\sum_S H_S$. Here $S\subseteq \{1,\ldots,n\}$ runs over all
subsets of qubits of cardinality $k$ and $H_S$ may be an arbitrary
Hermitian operator on $S$ tensored with the identity on all qubits
from $\{1,\ldots,n\}\backslash S$.
The locality of interactions in the definition above can be regarded as an
{\it algebraic locality}. It
should not be confused with
a {\it geometric locality} which can be defined only if
the set of qubits is endowed with a metric or a graph structure.
  A natural unit of energy set by
$H$ is given by the maximum operator norm of the interactions,
$J=\max_S ||H_S||$.  Let $\lambda(H)$ be the smallest eigenvalue of
$H$, i.e. the ground-state energy. Suppose we are promised that
either $\lambda(H)\le 0$ or $\lambda(H) \ge \delta$, where $\delta$
is at least $\frac{J}{{\rm poly}(n)}$. The Local Hamiltonian Problem
is formulated as a decision problem: given the data
$(n,\left\{H_S\right\},\delta)$, one has to decide whether
$\lambda(H)\le 0$.  A more formal definition is given in
Section~\ref{sec:LHP}. We will refer to the Local Hamiltonian
problem as LH-MIN indicating that it is the problem of estimating
the minimum eigenvalue of $H$.

If one considers a generic spin Hamiltonian $H$ that lacks any
additional structure except for the locality of interactions, it is
extremely unlikely that LH-MIN can be solved in polynomial time
(even on a quantum computer).  Indeed, it was shown by
Kitaev~\cite{KSV:computation} that LH-MIN is a complete problem in
the complexity class \QMA{} --- the quantum analogue of \NP. This
\QMA-completeness result applies even to Hamiltonians with $2$-qubit
nearest-neighbor interactions on 2D square
lattice~\cite{KKR:hamsiam,OT:qma}. Therefore, instead of looking for
efficient algorithms for evaluating the ground-state energy, we have
to focus on efficient {\em proving protocols} by which the prover (a
party with unlimited computational power) can prove an upper bound
on the ground-state energy to the verifier (a party that has
polynomial resources).

By definition, the inclusion $\mbox{LH-MIN}\in \QMA$ means that the
upper bound $\lambda(H)\le 0$ has an efficient quantum proving
protocol with one round of communication between the prover and the
verifier, see~\cite{watrous:ma}. One of the goals of the present
paper is to argue that there exists a large subclass of quantum
local Hamiltonians for which LH-MIN has an efficient {\it classical}
proving protocol with a constant number of communication rounds.
This subclass involves all local spin Hamiltonians whose matrix
elements in the standard basis of $n$ qubits satisfy the condition that {\it
all off-diagonal matrix elements are real and non-positive}. A nice
property of such Hamiltonians is that the corresponding Gibbs
density matrix $\rho=e^{-\beta \, H}/\tr{\left( e^{-\beta\,
H}\right)}$ has non-negative matrix elements in the standard basis
for any $\beta\ge 0$. From this non-negativity property of the Gibbs
matrix it follows by simple linear algebra arguments that the
ground-state $\ket{\Psi_0}$ of $H$ has non-negative real
coefficients, i.e. $\ket{\Psi_0}=\sum_i \alpha_i \ket{i}$ where
$\alpha_i \geq 0$. Thus one can associate a probability distribution
with the ground-state, $\mathbb{P}(i)=\frac{\alpha_i}{\sum_i
\alpha_i}$. If one is able to sample efficiently from this
distribution one can determine $\lambda(H)$ (for details see Section
\ref{sec:POSTBPP}). Because of the relation to stochastic processes
we have adopted the term {\em stoquastic} to refer to these
Hamiltonians. In the `standard' basis, these Hamiltonians have
non-positive off-diagonal matrix elements. This standard basis for
local Hamiltonians is typically the local spin-z basis, but one can
of course allow for local unitary basis changes without changing the
complexity of the problem.

Clearly, any classical spin Hamiltonian, i.e. a Hamiltonian which is
diagonal in the standard basis, falls into the stoquastic class.
Here are some 1-local and 2-local stoquastic operator on qubits:
\[
-X,
 \quad
 -X\otimes\, |z\ra\la z| \quad \mbox{for}\quad z \in \{0,1\},
\quad
-p\,X\otimes\, X -
q\, Y\otimes \, Y \quad \mbox{for any} \quad  0\le q\le p.
\]
It can be shown that all 2-local stoquastic Hamiltonians on qubits
can be generated by taking convex linear combinations of these
stoquastic 2-local interactions and all classical 2-local
interactions (composed solely from tensor products of $Z$)
\footnote{It can be shown that there are 3-local Hamiltonians on
qubits which are stoquastic, but not {\em termwise} stoquastic, i.e.
they cannot be written as a sum over stoquastic terms that acts on 3
qubits at the time.}.

Stoquastic Hamiltonians are very common in physics. Among spin-1/2
models, the well-studied ferromagnetic Heisenberg models and the
quantum transverse Ising model (considered for example by
Farhi~\cite{farhi+:adia_science} in the context of adiabatic quantum
computation) are stoquastic.  Another example is a Heisenberg
anti-ferromagnet on a cubic lattice (or more generally, on a
bipartite graph):
\[
H=\sum_{(j,k)} X_j\otimes X_k + Y_j\otimes
Y_k + Z_j\otimes Z_k.
\]
Here the qubits live at vertices of the lattice and the interactions
couple nearest-neighbors on the lattice. Although $H$ is not
directly stoquastic, it can be simply made so by a local change of
basis. Indeed, if a lattice admits a bi-coloring, one can apply $Z$
to every white vertex to flip the sign of $X\otimes X$ and $Y\otimes
Y$\footnote{The new basis coincides with the original one up to
phases of the basis vectors.}. This produces a stoquastic
Hamiltonian.

Although in this paper we focus only on spin-$1/2$ Hamiltonians, the
stoquastic class naturally extends to systems of qudits, or even
infinite-dimensional particles (e.g. harmonic oscillators). For
example, a system of spin-less interacting bosons is described (in
the first quantization formalism) by a Hamiltonian $H=K+U$, where
$K=-\frac{1}{2m}\sum_a \Delta_a $ is a kinetic energy (when the
vector potential is zero) and $U$ is a potential energy.
Off-diagonal matrix elements of $H$ come only from $K$. The
discretized version of the Laplacian,
$\Delta_a=\frac{d^2}{dx^2_a}=\sum_j |j+1\ra\la j| + |j\ra\la j+1| -
2|j\ra\la j|$ shows that all off-diagonal matrix elements of $K$ are
non-positive.  Outstanding examples in this category are bosonic
Bose-Einstein condensates and Helium-4~\cite{Cep}; there is a
general belief in the computational physics community that the
ground-state properties of such systems are ``easy" to simulate,
although no rigorous basis for this opinion seems to exist
presently.

All Josephson-junction qubit systems of the `flux'-type are
stoquastic.  The quantum-mechanics of any such system is that of a
collection of {\em distinguishable} (rather than bosonic or
fermionic) particles with a Hamiltonian $K+U$ as just
discussed~\cite{BKD}.  It was this observation that initiated the
present investigation, and indicated that flux qubits would not be
the most general choice for implementing adiabatic quantum
computation.

Other stoquastic Hamiltonians are identified by noting that bosonic
creation/annihilation operators $\hat{a}\, |j\ra=\sqrt{j}\, |j-1\ra$
and $\hat{a}^\dag\, |j\ra = \sqrt{j+1}\, |j+1\ra$ have non-negative
matrix elements in the occupation number basis. Therefore a hopping
operator $-\hat{a}^\dag_j \, \hat{a}_k - \hat{a}^\dag_k \,
\hat{a}_j$, and the entire class of bosonic Hubbard models, belongs
to the stoquastic class.  Among systems involving both spin-1/2 and
bosonic degrees of freedom, the Jaynes-Cummings model~\cite{WalMil},
and the spin-boson model~\cite{Legetal}, are also stoquastic when
suitable phases are associated with the vectors in the standard
basis.

Naturally, not {\em all} Hamiltonians in physics are stoquastic.
Many fermionic systems are non-stoquastic; the antisymmetry of the
(first-quantized) wavefunction causes it to have sign changes in the
position basis.  In the occupation-number (second-quantized) basis,
terms of both signs typically occur as off-diagonal matrix elements
on account of the anticommutation relations of the creation and
annihilation operators.  Special fermionic systems, like the spin
systems mentioned above, can avoid this `sign problem' but generic
fermionic systems do not. Hamiltonians of charged (bosonic or
fermionic) particles in the presence of a magnetic field will also
not be stoquastic (the Hamiltonian, and the ground-state are
typically complex).

Stoquastic Hamiltonians have also featured in recent work in quantum
information theory. In Ref. \cite{AT:adia} they are used to define
an adiabatic path algorithm that is derived from a classical
reversible Markov chain and in Refs.~\cite{Henley04,VMPC:peps} they are
similarly defined on the basis of a Monte-Carlo process that
generates the equilibrium distribution of some classical
Hamiltonian. In these constructions, there is a direct connection
between the rapid convergence of the Markov chain and the gap of the
resulting stoquastic Hamiltonian. In some sense these constructions,
and our results, are rigorous expressions and examples of the
physics folklore theorem which says that one can map ground-state
problems of $d$-dimensional Hamiltonians onto classical statistical
problems in $d+1$-dimensions \cite{suzuki}. In this paper we show in
fact that if some rigorously defined version of this folklore
statement were true than it would have the complexity-theoretic
consequence that $\QMA \subseteq \AM$, which we consider unlikely.
Thus as it stands, it is only the class of stoquastic Hamiltonians
that allow for this quantum-to-classical mapping.

\section{Summary of Main Results}
\label{sec:results}

Let us review our main results. Obviously, restricting ourselves to
a subclass of local Hamiltonians can only reduce the complexity of
LH-MIN which means that stoquastic LH-MIN belongs to the class \QMA.
On the other hand, stoquastic LH-MIN is \NP-hard, since it includes
all classical local Hamiltonians. Indeed, it was proved by
Barahona~\cite{barahona:np} that finding the ground-state energy of
the Ising model on the 3D cubic lattice with couplings $J \in
\{-1,0,+1\}$ is a \NP-complete problem.

Firstly, we prove that stoquastic LH-MIN belongs to the complexity
class \AM{}. \AM{} is a probabilistic analogue of \NP{} with two
rounds of communication between the prover and the verifier, see
Section~\ref{sec:AM}. The proof proceeds by mapping stoquastic
LH-MIN to the Approximate Set Size problem. We consider a
``partition function'' $Z=\tr{(G^L)}$, where $G=I-\beta\, H$ is a
non-negative matrix whose largest eigenvalue is $\mu=1-\beta\,
\lambda(H)$. If $L$ is a sufficiently large, $Z\approx \mu^L$ and
thus $Z$ provides enough information about $\lambda(H)$. Then we
convert $G$ into a sum of $0,1$-matrices thus expressing $Z$ as a
sum of a Boolean function over all input arguments. Evaluating this
sum is equivalent to the Approximate Set Size problem. The latter
problem admits a two-round interactive proof based on Carter-Wegman
universal hashing, see~\cite{GS:coin,CW:hash}. It should be noted
that \AM{} also contains a generalization of stoquastic LH-MIN in
which $G$ may be an arbitrary non-negative matrix specified by a
black box. In a sequel to this paper \cite{BBT:sma} we will
strengthen this result and prove that stoquastic LH-MIN is in a
class called SBP.

Secondly, we show that the $6$-local stoquastic Hamiltonian problem
is hard for the class \MA{} --- the probabilistic analogue of \NP
(see Section~\ref{sec:MA} for details). The main idea of the proof
is that any classical probabilistic machine can be simulated by a
classical circuit $C$ with {\it reversible} gates whose input
include ancillary random bits. Such a circuit can be transformed
into a coherent form $U_C$ by replacing each gate with a unitary
operator (which just permutes basis vectors) and replacing each
random bit with a coherent superposition $(|0\ra + |1\ra)/\sqrt{2}$.
Making use of the standard clock Hamiltonian
construction~\cite{KSV:computation} we can define a local
Hamiltonian $H$ whose ground-state energy is related to the maximum
acceptance probability of the quantum circuit $U_C$. The condition
that $U_C$ is composed only of classical gates guarantees that $H$
is an stoquastic Hamiltonian. We then prove that allowing Merlin to
feed quantum states into the verifying circuit does not give him any
additional cheating power as compared to the classical case.

Thirdly, we prove that for any constant $k$ $k$-local stoquastic
LH-MIN can be reduced in polynomial time to $2$-local stoquastic
LH-MIN. The proof is based on perturbation theory gadgets introduced
in~\cite{KKR:hamsiam}. We construct a new three-qubit gadget that
involves only stoquastic interactions, see Section~\ref{sec:gadgets}
for details.  A corollary of this result is that $2$-local
stoquastic LH-MIN is hard for \MA. The fact that the complexity of
$k$-local stoquastic LH-MIN does not depend upon $k$ indicates that
this problem might be complete for some well-defined computational
class, even though the nature of this class remains elusive to us.

Finally, we consider a special case of stoquastic LH-MIN in which
the Hamiltonian has a polynomial spectral gap (the difference
between the smallest and the second smallest eigenvalue is $1/{\rm
poly}(n)$ for some polynomial in $n$), see Section~\ref{sec:POSTBPP}
for details. In this case we prove that stoquastic LH-MIN belongs to
the class \POSTBPP{} --- the class of languages recognizable by
poly-time probabilistic Turing machines which produce the correct
answer (with constant error probability) conditioned on the value of
a `success flag' bit (the success probability may be exponentially
small though). The proof relies on the ideas borrowed from the
Green's Function Quantum Monte Carlo method,
see~\cite{Buonaura:GFMC} and we show that post-selected classical
computation gives us the power to sample from the ground-state
distribution.  This last result also implies that any decision
problem solved by an adiabatic quantum algorithm that uses only
stoquastic Hamiltonians is contained in \POSTBPP{}.

\subsection{Definition of the Local Hamiltonian Problem}
\label{sec:LHP}

We shall denote the smallest eigenvalue of a Hamiltonian $H$ by $\lambda(H)$.
\begin{defi}
\label{def:LHP1} For any integer $k$, polynomials $p_1(n)$ and
$p_2(n)$ define a set $\Omega(k,p_1,p_2)$ involving all $k$-local
$n$-qubit Hamiltonians $H=\sum_S H_S$ such that for any fixed $k \le
n<\infty$ one has
\begin{itemize}
\item $||H_S||\le p_1(n)$ for all subsets $S\subseteq \{1,\ldots,n\}$, $|S|=k$
\item Either $\lambda(H)\le 0$ or $\lambda(H)\ge 1/p_2(n)$
\end{itemize}
\end{defi}
Suppose we are given a Hamiltonian $H\in \Omega(k,p_1,p_2)$ and our
goal is to decide whether $\lambda(H)\le 0$. Clearly, the correct
decision can be made even if the interactions $H_S$ are specified
only up to some precision $\delta$ polynomial in $1/n$. Indeed, if
Hamiltonians $H$ and $H'$ are $\epsilon$-close in the operator norm,
$||H-H'||< \epsilon$, then their ground-state energies are also
$\epsilon$-close, $|\lambda(H)-\lambda(H')|<\epsilon$ (see for
example \cite{book:horn&johnson:matrix}). Thus, although
$\Omega(k,p_1,p_2)$ is a continuum set, we can safely assume that
any $H\in \Omega(k,p_1,p_2)$ is described by ${\rm poly}(n)$ bits.
In that sense we can regard $\Omega(k,p_1,p_2)$ as a set of finite
binary strings.

\begin{defi}[Local Hamiltonian Problem (LH-MIN)]
\label{def:LHP2} Given a description of a Hamiltonian $H\in
\Omega(k,p_1,p_2)$, decide whether $\lambda(H)\le 0$.
\end{defi}

\section{Stoquastic LH-MIN in \AM}
\label{sec:AM}

The complexity class \AM{} was introduced originally by
Babai~\cite{babai} as a class of decision problems that possess a
randomized interactive proof with two-way communication between the
prover (Merlin) having unlimited computational resources and the
verifier (Arthur) capable of doing only polynomial-time computation.
It is a remarkable property of the class \AM{} that any proving
protocol with constant number of communication rounds\footnote{A
communication round involves a single message sent from one party to
the other.} can be simulated by a protocol with just two
rounds~\cite{babai}, such that the first message is sent from Arthur
to Merlin, and the second one backwards.


We shall mostly consider promise problems. Let $\Sigma=\{0,1\}$ and
let $\Sigma^n$ be a set of $n$-bit strings and $\Sigma^*$ be a set
of all finite binary strings. A promise problem can be regarded as
a pair of non-overlapping subsets of binary strings $L_{\yes},L_{\no}\subseteq \Sigma^*$
corresponding to
positive and negative instances. An Arthur-Merlin proving protocol
for a membership $x\in L_{\yes}$ involves Arthur's question $q\in
\Sigma^{p(|x|)}$ and Merlin's response $r\in \Sigma^{p(|x|)}$, where
$p$ is a fixed polynomial and $|x|$ is the number of bits in $x$.
Arthur's question is just a random bit string drawn from the uniform
distribution. The response $r$ may be an arbitrary function of $x$
and $q$. Once the communication is completed, Arthur has at his
disposal all the data $x,q,r$.  Then he runs a \BPP{} test
$V(x,q,r)$  that outputs either $1$ (accept the proof) or $0$
(reject the proof).

A proving protocol must obey
soundness and completeness properties.
Completeness means that for positive instances, $x\in L_{yes}$,
Merlin has a strategy (i.e. a response functions $r(x,q)$)
for which Arthur's acceptance probability is close to $1$.
Soundness means that for negative instances, $x\in L_{\no}$,
Arthur's acceptance probability is close to $0$ for all
possible Merlin's strategies. Here is a formal definition\footnote{It is
known that completeness with a constant error probability is equivalent to
perfect completeness, see
~\cite{furer89completeness}.}:

\begin{defi}
A promise problem  $L=L_{\yes}\cup L_{\no}\subseteq \Sigma^*$
belongs to the class \AM{} iff there exists a polynomial $p$ and a
\BPP{} predicate $V(x,q,r)$ defined for any $q,r\in
\Sigma^{p(|x|)}$, such that \bea x\in L_{\yes} &\implies & \exists
\, r(x,q) \,  \,
\prob{\left[V(x,q,r(x,q))=1\right]}\ge 2/3 \nn \\
x\in L_{\no} &\implies & \forall\, r(x,q)\,\,
\prob{\left[V(x,q,r(x,q))=1\right]}\le 1/3 \eea where $q\in
\Sigma^{p(|x|)}$ is a uniformly distributed random bit string.
\end{defi}

The main goal of this section is to show that LH-MIN for stoquastic
Hamiltonians belongs to the class \AM. Moreover, we will prove that
evaluation of the largest eigenvalue of any $n$-qubit non-negative
matrix whose matrix elements are efficiently computable is a problem
that naturally sits in \AM. This result applies even to matrices
that lack any additional structure like locality or sparseness. To
emphasize this point, we will formulate all results in terms of {\it
black box matrices}. A black box matrix $G$ of size $2^n\times 2^n$
is an oracle that takes as input two binary strings $x,y\in
\Sigma^n$ and returns a matrix element $G_{x,y}=\la x|G|y\ra$
written in the binary form. We shall always assume that any matrix
element $G_{x,y}$ has at most ${\rm poly}(n)$ binary digits (see the
remark after Definition~\ref{def:LHP1}). In the case when $G$ is
specified by a local Hamiltonian, there is no need to query the
oracle, since $G$ has a concise representation and we can compute
$G_{x,y}$ in a time ${\rm poly}(n)$.

Let $G$ be a black box non-negative matrix and let $\mu(G)$ be the
largest eigenvalue of $G$. To cast the evaluation of $\mu(G)$ into a
decision problem we shall introduce two thresholds: an {\it upper
threshold} $\mu_+$ and a {\it lower threshold} $\mu_-$, such that
$0<\mu_-<\mu_+$ and the separation between $\mu_-$ and $\mu_+$ is
large enough.

\begin{defi}
\label{def:Lambda}
For any polynomial $p(n)$ define a set $\Lambda(p)$ consisting of all
$4$-tuples $(n,G,\mu_+,\mu_-)$ such that
$n$ is an integer $1\le n<\infty$, $\mu_\pm$ are positive numbers
such that
$\log(\mu_+) - \log(\mu_-) \ge 1/p(n)$,
and $G$ is a $2^n\times 2^n$ real symmetric matrix such that
\begin{itemize}
\item $0\le G_{x,y}\le 1$ for all $x,y\in \Sigma^n$.
\item Either $\mu(G)\ge \mu_+$ or $\mu(G)\le \mu_-$.
\end{itemize}
\end{defi}


Suppose we are given a $4$-tuple $(n,G,\mu_+,\mu_-)\in \Lambda(p)$
and our goal is to decide whether $\mu(G)\ge \mu_+$. According to
the Weyl perturbation theorem (see the remark after
Definition~\ref{def:LHP1}), the correct decision can be made even if
the matrix elements $G_{x,y}$ and the numbers $\mu_\pm$ are
specified only up to some precision $\delta$ polynomial in $2^{-n}$.
Indeed, if $G$ and $G'$ are two $2^n\times 2^n$ matrices  such that
matrix elements of $G$ and $G'$ are $\epsilon$-close, then
$|\mu(G)-\mu(G')|\le ||G-G'||\le 2^n\epsilon$.
 Thus, although $\Lambda(p)$
is a continuum set, we can safely assume that the numbers $\mu_\pm$
and any matrix element $G_{x,y}$ are described by ${\rm poly}(n)$
bits.

\begin{defi}[Stoquastic Largest Eigenvalue Problem]
Given is a $4$-tuple $(n,G,\mu_+,\mu_-)\in \Lambda(p)$ where $G$ is
specified by a black box. Decide whether $\mu(G)\ge \mu_+$.
\end{defi}

{\it Remark:} One can easily see that stoquastic LH-MIN is a special
case of the problem above. Indeed, if $H\in \Omega(k,p_1,p_2)$ is a
$k$-local stoquastic Hamiltonian on $n$-qubits, see
Definitions~\ref{def:LHP1},\ref{def:LHP2}, one can define a
non-negative matrix $G=(1/2)(I-H/C)$,
where $C$ is an efficiently computable polynomial upper bound on the norm $\|H\|$,
for example, $C=\sum_S \|H_S\|$. Off-diagonal matrix elements of $G$ are non-negative
because $H$ is stoquastic. Diagonal matrix elements are non-negative because $I-H/C$
is a positive semi-definite operator. Since $\|G\|\le 1$, we conclude that
$0\le G_{x,y}\le 1$. One can also define the thresholds $\mu_+=1/2$ and
$\mu_-=(1/2)(1-1/C p_2(n))$. Clearly, the resulting $4$-tuple
$(n,G,\mu_+,\mu_-)\in \Lambda(p)$ for a proper choice of the
polynomial $p$.

\begin{theorem}
\label{thm:AM} Stoquastic Largest Eigenvalue Problem belongs to
the class \AM.
\end{theorem}

{\bf Proof:}
Consider any $4$-tuple $(n,G,\mu_+,\mu_-)\in \Lambda(p_1)$ where $p_1$ is a fixed polynomial.
Instead of proving the lower bound $\mu(G)\ge \mu_+$ Merlin will actually try to prove
a lower bound $\tr(G^L)\ge (\mu_+)^L$ where $L$ is a large even integer.
Note that
\bea
\mu(G) \ge \mu_+ &\implies & \tr(G^L)\ge \mu_+^L \nn \\ \nn \\
\mu(G) \le \mu_- &\implies & \tr(G^L) \le 2^n\, \mu_-^L. \nn
\eea
The separation between the value of the trace for positive and negative
instances is thus given by
\[
\frac{\tr(G^L)_{\yes}}{\tr(G^L)_{\no}} \ge 2^{\frac{L}{p_1(n)}\, - n}.
\]
If one chooses $L=2np_1(n)$, the separation is $2^n$.

The next step is to represent the evaluation of the trace $\tr(G^L)$
as a counting problem. As was mentioned after Definition~\ref{def:Lambda},
 we can assume that
the matrix elements $G_{x,y}$ have at most $p_2(n)$ digits, where
$p_2(n)$ is a polynomial. In order to define the counting problem,
we shall represent $G$ as an average over an ensemble of
$0,1$-matrices $G(t)$, where $t$ is a random uniformly
distributed binary string $t\in \Sigma^{p_2(n)}$, that is \be
\label{G(t)} G=\frac1{2^m} \sum_{t\in \Sigma^{m}} G(t), \quad
m\equiv p_2(n). \ee Any member of the ensemble $G(t)$ is a binary
matrix, that is, matrix elements of $G(t)$ take only values $0$ and
$1$. This representation is efficient in the sense that for any
fixed strings $x,y,t$ one can find a matrix element $\la
x|G(t)|y\ra$ by making one query to the black box for $G$ and
performing a polynomial-time computation. Details of the
representation Eq.~(\ref{G(t)}) are not essential for the analysis
of the proving protocol, so we postpone its proof until
 Lemma~\ref{lemma:G(t)}.
Now we have
\[
\tr(G^L)=\frac1{2^{m\, L}} \sum_{t_1,\ldots,t_L} \tr{(G(t_1)\cdots
G(t_L))} \equiv \frac1{2^{m\, L}} \sum_s F(s),
\]
where $s=(t_1,\ldots,t_L,x_1,\ldots,x_L)$ is a binary string of length $(m+n)L$
and $F(s)$ is a Boolean function
\[
F(s)=\la x_1|G(t_1)|x_2\ra\, \la x_2|G(t_2)|x_3\ra \cdots
\la x_L|G(t_L)|x_1\ra \in \{0,1\}.
\]
Evaluation of $F(s)$
requires $L$ black box queries and
polynomial-time computation.
Summarizing, the value of $\tr(G^L)$ is proportional to a cardinality of
a set $\Omega\subseteq \Sigma^{(m+n)L}$ supporting the function $F$,
\[
\tr{(G^L)}=\frac1{2^{mL}} \, |\Omega|, \quad
\Omega=\{ s\in \Sigma^{(m+n)L}\, : \, F(s)=1\},
\]
and membership $s\in \Omega$ can be efficiently verified. Note that
there is large enough separation between the cardinality of $\Omega$
for positive and negative instances: \bea
\mu(G) \ge \mu_+ & \implies & |\Omega|\ge  \lrg \nn  \\
\mu(G) \le \mu_- & \implies & |\Omega|< \sml, \nn
\eea
where
\be
\label{ls}
\lrg=2^{L(p_2(n)+\log \mu_+)}
\quad \mbox{and} \quad
\sml = 2^{L(p_2(n)+\log \mu_- + \frac{n}{L})},
\ee
such that
\be
\label{ls1}
\lrg = 2^n \cdot \sml \quad \mbox{if} \quad L= 2np_1(n).
\ee
Thus it suffices for Merlin to prove a lower bound $|\Omega|\ge \lrg$.

We can now invoke the Goldwasser and Sipser approximate counting
protocol~\cite{GS:coin} based on Carter-Wegman universal hashing
functions~\cite{CW:hash}. Recall that $\Omega$ is a set of $k$-bit
strings, where $k=L(n+p_2(n))$. The main idea of~\cite{GS:coin} is
that Arthur can compress $k$-bit strings to shorter $b$-bit strings
using randomly chosen linear hash functions. One can choose
parameters of the hashing such that the image $h(\Omega)\subseteq
\Sigma^b$ is sufficiently dense (for positive instances).
 Arthur estimates the volume of $h(\Omega)$
  using the standard Monte-Carlo method: he generates a
large list of random $b$-bit strings and estimates the fraction of
strings that belong to $h(\Omega)$. At this stage he needs Merlin's
help, since a membership in the set $h(\Omega)$ is no longer
efficiently verifiable because each string in $\Sigma^b$ may have
exponentially large number of pre-images.
On the other hand, Merlin can prove a membership in the set $h(\Omega)$
by sending Arthur any of pre-images.
 In Appendix \ref{app:a}
we give some details of the parameters of the hash functions.

Now we prove the Lemma underlying Eq. (\ref{G(t)})

\begin{lemma}
\label{lemma:G(t)}
Let $I_m=\{2^{-m}\, p\}_{p=0,\ldots,2^m-1}$ be the set of all real numbers
between $0$ and $1$ having at most $m$ binary digits.
Let $g\, : \, \Sigma^n \to I_m$ be a function specified by a black box.
Then there exists a Boolean function
$f\, : \, \Sigma^n\times \Sigma^m \to \Sigma$ such that
\[
g(x)=\frac1{2^m} \sum_{t\in \Sigma^m} f(x,t)
\quad \mbox{for all} \quad x\in \Sigma^n.
\]
Besides, $f(x,t)$ can be represented by a circuit of length $poly(n+m)$
making one query to the black box.
\end{lemma}
\begin{proof}
Let $d_j(x)$ be the $j$-th binary digit of $g(x)$, that is
\[
g(x)=\sum_{j=1}^m \frac1{2^j}\, d_j(x).
\]
Define $m$ auxiliary Boolean functions
\bea
f_1(x,t) &=& d_1(x) \wedge t_1, \nn \\
f_2(x,t) &=& d_2(x) \wedge (\neg\, t_1) \wedge t_2, \nn \\
f_3(x,t) &=& d_3(x) \wedge (\neg \, t_1)\wedge (\neg \, t_2) \wedge t_3, \nn \\
&\cdots & \nn \\
f_m(x,t) &=& d_m(x) \wedge (\neg \, t_1)\wedge \ldots \wedge (\neg\, t_{m-1}) \wedge t_m. \nn
\eea
Here $t_j$ is the $j$-th bit of $t$.
Clearly,
\[
\frac1{2^m} \sum_{t\in \Sigma^m} f_j(x,t) = \frac1{2^j} \, d_j(x),
\quad j=1,\ldots,m.
\]
By definition, the functions $f_j$ and $f_k$ are mutually exclusive for $j\ne k$.
Therefore
\[
\sum_{j=1}^m f_j=f_1\vee f_2\vee \ldots \vee f_m.
\]
Thus we can define the desired function $f(x,t)$ as
$f=f_1\vee f_2\vee \ldots \vee f_m$.
\end{proof}

{\it Comment:} The representation Eq.~(\ref{G(t)}) corresponds to choosing
$g(x)=\la y|G|z\ra$, where $x$ is a concatenation of the strings $y$ and $z$.

\section{Stoquastic LH-MIN is \MA-hard}
\label{sec:MA}

In order to show that stoquastic LH-MIN is \MA-hard we will view
Arthur's \BPP{} circuit as a quantum circuit. This quantum circuit
will take as input: a quantum state $\ket{\xi}$ from Merlin, a set
of $\ket{+}$ states (to simulate randomness) and some ancillas set
to $\ket{0}$. The quantum circuit consists only of classical
reversible gates and at the end Arthur measures a single qubit
$q_{\rm out}$ in the z-basis. He obtains 1 with high probability if
the answer to his decision problem is yes; otherwise he obtains 0
with high probability. If Merlin can only provide a classical state
it is clear that the class of decision problems that can be solved
this way is equal to \MA. Before we argue that this new class of
decision problems is equal to \MA, let us give the proper
definition.

\begin{defi}[${\rm MA_q}$]
A promise problem  $L_{\yes},L_{\no}\subseteq \Sigma^*$
belongs to the class ${\rm MA_q}$ iff there exists a polynomial $p$
and a classical reversible circuit $V_x$ that takes an input in
$(\mathbb{C}^2)^{\otimes p(|x|)}$ and is followed by a single qubit
measurement, such that \bea x\in L_{\yes} &\implies & \exists \,
\ket{\xi} \,  \,
\prob{\left[V_x(\ket{00 \ldots 0},\ket{+}^{\otimes r},\ket{\xi})=1\right]}\ge 2/3 \nn \\
x\in L_{\no} &\implies & \forall\, \ket{\xi} \, \,
\prob{\left[V_x(\ket{00 \ldots 0},\ket{+}^{\otimes
r},\ket{\xi})=1\right]}\le 1/3. \eea
\end{defi}

\begin{lemma}
${\rm MA}={\rm MA}_q$.
\end{lemma}

\begin{proof}
${\rm MA}_q \subseteq {\rm MA}$: Let $(L_{\yes},L_{\no})$
be a promise problem in ${\rm MA}_q$. If $x \in
L_{\yes}$ we have $\mathbb{P}(V_x(\ket{+}^{\otimes r},\ket{00\ldots 0},
\ket{\xi})=1) \geq 2/3$. Let $\Pi_1=\ketbra{1}_{q_{\rm out}}$. We
can write the success probability as \be \mathbb{P}(1)=\bra{\xi} M
\ket{\xi} \geq 2/3, \ee where $M=\bra{00 \ldots 0, +^{\otimes r}}
V_x^{T} \Pi_1 V_x \ket{00 \ldots 0,+^{\otimes r}}$. We note that the
observable $M$ is diagonal in the standard basis, i.e.
$M=\frac{1}{2^r} \sum_z a_z \ketbra{z}$ where $a_z$ is a
non-negative integer. This implies that $\lambda_{\rm
max}(M)=\max_{\xi} \bra{\xi} M \ket{\xi}$ is achieved for some bit
string $\ket{\xi}=z_{\rm max}$. Thus there exists a bit-string for
which $\mathbb{P}(1) \geq 2/3$ and this bit-string will be the input
for the {\rm MA}-verifier. If $x \in L_{\no}$, we have that $\forall
\xi\; \mathbb{P}(1)=\bra{\xi} M \ket{\xi} \leq 1/3$, thus
this also holds for the subset of all classical inputs from Merlin. \\
${\rm MA} \subseteq {\rm MA}_q$: let a decision problem be in \MA.
If $x \in L_{\yes}$, the classical witness can be used as input to the
${\rm MA_q}$-verifier and gives $\mathbb{P}(1)\geq 2/3$. If $x
\in L_{\no}$, we need to argue that Merlin cannot cheat by giving
Arthur a quantum state. Since the problem is in \MA, we have that
$\forall z\; \mathbb{P}(1)=\bra{z} M \ket{z} \leq 1/3$. Since $M$ is
diagonal in the $z$-basis, this implies that $\lambda_{\rm max}(M)
\leq 1/3$ and thus there is no quantum state with expectation value
higher than $1/3$ with respect to $M$.
\end{proof}

Since Arthur's verifying circuit in ${\rm MA}_q$ is a quantum
circuit, one can apply Kitaev's circuit-to-Hamiltonian construction
to ${\rm MA}_q$ and prove that the ground-state energy problem for a
6-local stoquastic Hamiltonian is ${\rm MA}_q$=${\rm MA}$-hard.

\begin{lemma}
6-local stoquastic {\rm LH-MIN} is ${\rm MA}$-hard.
\label{lemma:mahard}
\end{lemma}

\begin{proof}
Let $V_x$ be Arthur's verifying circuit that has an input of $r$
qubits in the state $\ket{+}$ (labeled as coin-qubits), $k$ ancilla
qubits in the state $\ket{00 \ldots 0}$ (labeled as anc-qubits) and
a quantum state $\ket{\xi}$ with $s$ qubits. Let $V_x$ have a total
of $T$ reversible classical gates, denoted as $R_T \ldots R_2 R_1$.
W.l.o.g. we can assume that each gate is a Toffoli gate, since these
gates are universal for classical reversible computation. We follow
the Hamiltonian construction in \cite{KSV:computation} (see also
\cite{AN:qnp}). Let $H^{(5)}=H_{\rm in}+H_{\rm out}+H_{\rm
prop}+H_{\rm clock}$ be a Hamiltonian acting on $T$ clock-qubits
labeled by $t=1 \ldots T$ and $n=r+k+s$ computational qubits. We
have \bea
H_{\rm in}& = & \sum_{i=1}^r \ket{-}\bra{-}_{{\rm coin},i} \otimes \ket{0}\bra{0}_{t=1}+\sum_{j=1}^k \ket{1}\bra{1}_{{\rm anc},j} \otimes \ket{0}\bra{0}_{t=1},\nonumber \\
H_{\rm out}& = & \ket{0}\bra{0}_{q_{\rm out}} \otimes \ket{1}\bra{1}_{t=T}, \nonumber \\
H_{\rm clock} & = & \sum_{t=1}^T \ket{01}\bra{01}_{t-1,t}.
\eea
Furthermore, $H_{\rm prop}=\sum_{t=1}^T H_{\rm prop}(t)$ with
\bea
H_{\rm evolv}(1)& =& \ketbra{00}_{1,2}+\ketbra{10}_{1,2}
-R_1 \otimes (\ket{10}\bra{00}_{1,2}+\ket{00}\bra{10}_{1,2}), \nonumber \\
H_{\rm evolv}(t)& = & \ketbra{100}_{t-1,t,t+1}+\ketbra{110}_{t-1,t,t+1} \nonumber \\
& & -R_t\otimes (\ket{110}\bra{100}_{t-1,t,t+1}+\ket{100}\bra{110}_{t-1,t,t+1}) ,\;\;1 < t < T \nonumber \\
H_{\rm evolv}(T)& =& \ketbra{10}_{T-1,T}+\ketbra{11}_{T-1,T}
-R_T\otimes (\ket{11}\bra{10}_{T-1,T}+\ket{10}\bra{11}_{T-1,T}).
\eea It was proved in \cite{KSV:computation} that if there exists a
$\ket{\xi}$ such that $V_x$ outputs 1 with probability larger than
or equal to $1-\epsilon$ then $\lambda(H^{(5)}) \leq \epsilon$. If
on the other hand for all $\ket{\xi}$ $V_x$ outputs 1 with
probability smaller or equal to $\epsilon$, then $\lambda(H^{(5)})
\geq \frac{c(1-\epsilon)}{T^3}$ for some constant $c$. Thus the
ground-state energy problem of this Hamiltonian is ${\rm
MA}_q$-hard. We need only to verify that this Hamiltonian $H^{(5)}$
is of the stoquastic-type. The only terms that are off-diagonal in
the computational basis can be found in $H_{\rm prop}$ and $H_{\rm
in}$. Inspection of these terms confirms that the Hamiltonian is
stoquastic.
\end{proof}

{\em Remarks}: One may wonder whether one can extend the class ${\rm
MA}_q$ to a class in which Arthur's verification circuit is more
quantum, while the corresponding Hamiltonian is still stoquastic.
One possibility is to allow for a measurement in the x-basis
(instead of the z-basis) at the end, see \cite{BBT:sma}.

\section{Perturbation Theory Gadgets for Stoquastic Hamiltonians}
\label{sec:gadgets}

The goal of this section is to understand whether the complexity of
stoquastic $k$-local LH-MIN depends upon $k$ --- the number of
qubits involved in the interactions. We will answer this question
for Hamiltonians that are termwise-stoquastic, i.e., those having a
decomposition $H=\sum_S H_S$, where $S$ runs over subsets of $k$
qubits and $H_S$ is a stoquastic Hamiltonian acting on the subset
$S$.  Direct inspection shows that all examples of stoquastic
Hamiltonians encountered in the paper are also termwise-stoquastic
and for 2-local Hamiltonians these notions coincide.

\begin{theorem}
Let $k$ be any constant. Any instance of $k$-local termwise
stoquastic LH-MIN can be reduced in polynomial time to $2$-local
stoquastic LH-MIN.
\end{theorem}

Throughout this section we will use the word stoquastic to refer to
Hamiltonians that are termwise-stoquastic. Our main technical tool
is the perturbation theory gadgets developed in~\cite{KKR:hamsiam}
and extended in~\cite{OT:qma}. The proof can be organized in three
parts. Firstly we reduce $k$-local interactions to $3$-local
interactions using a variant of the subdivision gadget
from~\cite{OT:qma}. This gadget only requires perturbation theory to
second-order.  The second step is to bring a stoquastic $3$-local
Hamiltonian into a special form \be\label{xxx} H=\helse
-\sum_{(j,k,l)} h_{jkl}\, X_j \, X_k \, X_l, \ee where $\helse$ is a
$2$-local stoquastic Hamiltonian, $(j,k,l)$ labels triples of
qubits, and $h_{jkl}$ are non-negative constants. We shall refer to
Hamiltonians having a decomposition as in Eq.~(\ref{xxx}) as {\it
triple-X $3$-local Hamiltonians}. In order to implement the second
step a new three-qubit gadget will be constructed. The final step is
to reduce $3$-qubit interactions $-h_{jkl}\, X_j \, X_k \, X_l$ to
$2$-local interactions.  This can be done using the three-qubit
gadget of ~\cite{KKR:hamsiam}. Throughout this section we follow the
notation of~\cite{KKR:hamsiam} and~\cite{OT:qma}.

\subsection{Reduction to $3$-local interactions: the subdivision gadget}
Using the standard operator algebra basis of $n$ qubits, any
stoquastic $k$-local Hamiltonian $\htar$ can be written as
\[
\htar=\Omega\, I - \sum_{(j_1,\ldots,j_k)} \sum_{a_1,\ldots,a_k}
h_{j_1,\ldots,j_k}^{\alpha_1,\ldots,\alpha_k}\,
E_{j_1}^{\alpha_1} E_{j_2}^{\alpha_2} \cdots E_{j_k}^{\alpha_k} +
{\rm h.c.}
\]
Here $(j_1,\ldots,j_k)$ labels subsets of $k$ qubits, $\alpha$
labels one-qubit matrices $E^0=|0\ra\la 0|$, $E^1=|0\ra\la 1|$,
$E^2=|1\ra\la 0|$, $E^3=|1\ra\la 1|$, and
$h_{j_1,\ldots,j_k}^{\alpha_1,\ldots,\alpha_k}$ are non-negative
constants. The energy shift $\Omega\, I$ is introduced in order to
make all diagonal matrix elements of $\htar$ non-positive. Let us
partition each subset $(j_1,\ldots,j_k)$ into two non-overlapping
subsets of nearly equal size. Then we can rewrite $\htar$ as
\[
\htar=\Omega\, I - \sum_{a=1}^M ( C_a \otimes D_a + C_a^\dag
 \otimes D_a^\dag), \quad M=4^k {n \choose k}
\]
where $C_a$ and $D_a$ are operators having the following properties:\\
(1) All $C_a$ and $D_a$ have non-negative matrix elements,\\
(2) $C_a$ and $D_a$ act on non-overlapping subsets of
at most $\lceil k/2\rceil$ qubits, \\
(3) $C_a^\dag C_a$ and $D_a D_a^\dag$ are diagonal.\\
Since we regard $k$ as a constant, the number of terms in the sum is
polynomial, $M={\rm poly}(n)$.

Let us introduce $M$ mediator qubits and consider a
Hamiltonian $\tilde{H}$ acting on $n$ data qubits and $M$ mediator qubits:
\[
\tilde{H}=H+V,
\quad
H=\Delta \sum_{a=1}^M I_{\rm data} \otimes |1\ra\la 1|_a,
\quad
V=-\sqrt{\Delta} \sum_{a=1}^M (C_a + D_a^\dag)\otimes \sigma^+_a + (C_a^\dag + D_a)\otimes \sigma^-_a
+Q\otimes I_M,
\]
where $Q=\sum_{a=1}^M (C_a^\dag C_a + D_a D_a^\dag)$, $\sigma^+
=|1\ra\la 0|$, $\sigma^-=|0\ra\la 1|$. As for $\Delta$, it must be
chosen such that $||V||\ll \Delta$. Note that all terms in $H$ and
$V$ are stoquastic. Denote the Hilbert space of $n$ data qubits as
$\calH_{\rm data}$. Then $H$ has zero-energy levels defining the
eigen-subspace $\calL_-=\calH_{\rm data}\otimes |0^{\otimes M}\ra$
separated from the rest of the spectrum by a gap $\Delta$.
Considering $V$ as a perturbation, we compute the self-energy
operator \be\label{self_energy} \Sigma_-=V_{--} + V_{-+} G_+ V_{+-}
+ V_{-+} G_+ V_{++} G_+ V_{+-} + V_{-+} G_+ V_{++} G_+ V_{++} G_+
V_{+-} + \cdots \ee up to second-order of the perturbation theory
one gets\footnote{To avoid a proliferation of ${\rm poly}(n)$
bounds, we treat all terms proportional to $||C_a||$, $||D_a||$, or
$M$ as $O(1)$. In general all these terms can be bounded by ${\rm
poly}(n))$. Since we are free to choose $\Delta$ polynomially large,
the bounds $O(\Delta^{-1/2})$ and $O({\rm poly}(n)\Delta^{-1/2})$
are equally good.}
\[
\Sigma_-(z)=- \sum_{a=1}^M ( C_a \otimes D_a + C_a^\dag \otimes
D_a^\dag) + O(\Delta^{-1/2}) \quad \mbox{for any} \quad z=O(1).
\]
Accordingly, the ground-state energy of $\tilde{H}$ approximates the
ground-state energy of $\htar-\Omega I$ with precision
$\delta=O(\Delta^{-1/2})$.  This reduces $k$-local stoquastic LH-MIN
to $\lceil k/2\rceil+1$-local stoquastic LH-MIN. By repeating this
reduction $O(\log{(k)})$ times\footnote{After each iteration we have
to introduce an energy shift $\Omega\, I$, since the terms $C_a^\dag
C_a + D_a D_a^\dag$ may produce positive matrix elements on the
diagonal.} we end up with a $3$-local stoquastic Hamiltonian.

For obvious reasons the subdivision gadget cannot transform
$3$-local terms into $2$-local terms.  However, we can use it to
reduce the variety of $3$-local terms which have to be dealt with
using different (and more complicated) methods.

If one considers all possible  $3$-local terms proportional to
$-E^{\alpha_1}_1 E^{\alpha_2}_2 E^{\alpha_2}_2$, there are
essentially four different types of terms (up to permutations of
qubits and bit flips $0\leftrightarrow 1$) shown in the first column
of Table~1. By choosing the operators $C_a$ and $D_a$ from the
second and the third column, one can reduce interactions of type (a)
to type (b), type (b) to type (c), and finally type (c) to type (d).
This requires at most three repetitions of the subdivision gadget.
Now we can assume that a Hamiltonian has the form
\be\label{reduced_3_local} \htar=\helse - \sum_{(j,k,l)}
\sum_{\alpha,\beta,\gamma=\pm } h^{\alpha,\beta,\gamma}_{j,k,l} \,
\sigma_j^{\alpha} \otimes \sigma_k^{\beta} \otimes
\sigma_l^{\gamma}, \ee where $\helse$ is a stoquastic $2$-local
Hamiltonian, $(j,k,l)$ labels triples of qubits, and
$h^{\alpha,\beta,\gamma}_{j,k,l}\ge 0$.

\begin{table}
\centerline{\begin{tabular}{|c|l|c|c|}
\hline
& $3$-local term & choice of $C_a$ & choice of $D_a$ \\
\hline
(a) & $-|000\ra\la 000|_{jkl}$ & $|00\ra\la 00|_{jk}$ & $|0\ra\la 0|_l$ \\
\hline
(b) & $-|000\ra\la 100|_{jkl}$ & $|00\ra\la 10|_{jk}$ & $|0\ra\la 0|_l$ \\
\hline
(c) & $-|000\ra\la 110|_{jkl}$ & $|00\ra\la 11|_{jk}$ & $|0\ra\la 0|_l$ \\
\hline
(d) & $-|000\ra\la 111|$ &  &  \\
\hline
\end{tabular}}
\caption{Successive application of the subdivision gadget with the
choice of $C_a$ and $D_a$ as above reduces any $3$-local term to a
term of type (d).}
\end{table}

\subsection{Reduction from $3$-local to special $3$-local stoquastic Hamiltonians}

Our next goal is to construct a gadget reducing the stoquastic
Hamiltonian of Eq.~(\ref{reduced_3_local}) to a special $3$-local
Hamiltonian, see Eq.~(\ref{xxx}). To simplify the discussion, let us
first consider a Hamiltonian Eq.~(\ref{reduced_3_local}) with a
single $3$-local term:
\[
\htar=\helse-3(B_1 \otimes B_2 \otimes B_3 + B_1^\dag \otimes
B_2^\dag \otimes B_3^\dag), \quad B_j=\sigma^+ \quad \mbox{or} \quad
B_j=\sigma^-.
\]
where $\helse$ is a $2$-local stoquastic Hamiltonian, and the
factor $3$ is introduced for convenience.  We shall need three
mediator qubits which will be labeled by $1,2,3$. Consider a
Hamiltonian $\tilde{H}$ acting on $n$ data qubits and three
mediator qubits:
\begin{eqnarray}
\tilde{H} &=& H+V, \quad H=I_{\rm data}\otimes H_M  \nonumber \\
H_M&=&-\frac12 \Delta_x \, (X_1 \otimes X_2 \otimes X_3 - I)
-\frac14{\Delta_z}\, (Z_1 \otimes Z_2  + Z_2 \otimes Z_3 + Z_1
\otimes Z_3 - 3\, I), \nonumber  \\
V &=& -\omega\, \sum_{j=1}^3 B_j\otimes \sigma^+_j + B_j^\dag
\otimes \sigma^-_j + \helse\otimes I_M. \label{HV}
\end{eqnarray}

The parameters $\omega,\Delta_x,\Delta_z$ must be chosen as
\be\label{scaling} \omega=\delta^{-4}, \quad \Delta_x=\delta^{-5},
\quad \Delta_z=\delta^{-6}, \quad 0<\delta\ll 1. \ee It will be
shown later that $\delta$ is the precision up to which the
ground-state energy of $\tilde{H}$ approximates the ground-state
energy of $\htar$ (as before, we assume for simplicity that
$||B_j||$ and $||\helse||$ are of order $O(1)$). Note that
Eq.~(\ref{scaling}) implies $\omega\ll \Delta_x\ll \Delta_z$. Also
note that all local terms in $H$ and $V$ are stoquastic. The only
$3$-local term in $\tilde{H}$ is the one proportional to $-X_1 X_2
X_3$, so that $\tilde{H}$ is a special $3$-local stoquastic
Hamiltonian.

The Hamiltonian $H_M$ is diagonal in the basis of states
\be\label{eigen_basis} |\Psi^\pm\ra = \frac1{\sqrt{2}}\, |000\ra \pm
\frac1{\sqrt{2}}\, |111\ra, \quad \mbox{and} \quad |\phi^\pm_j\ra =
X_j \, |\Psi^\pm\ra, \quad j=1,2,3. \ee The spectrum of $H_M$ is
illustrated in Figure~1. By construction, $H_M$ has a unique
ground-state $|\Psi^+\ra$ having zero energy\footnote{One should not
confuse labels $\pm$ of the states $\Psi^\pm$ and $\phi^\pm$ with
the labels $\pm$ referring to the low-energy and high-energy
subspaces that appear in the perturbative series
Eq.~(\ref{self_energy}).}, while the first excited state
$|\Psi^-\ra$ has energy $\Delta_x$. The top part of the spectrum
involves six nearly-degenerate (as long as $\Delta_x\ll \Delta_z$)
states $\phi^\pm_j$. Since $||V||=O(\omega)\ll \Delta_x$, we can
treat $V$ as a perturbation and compute the self-energy operator on
the zero-energy subspace of $H$, that is $\calL_-=\calH_{\rm
data}\otimes |\Psi^+\ra$.

\begin{figure}
\label{fig:transitions}
\centerline{\includegraphics[height=5cm]{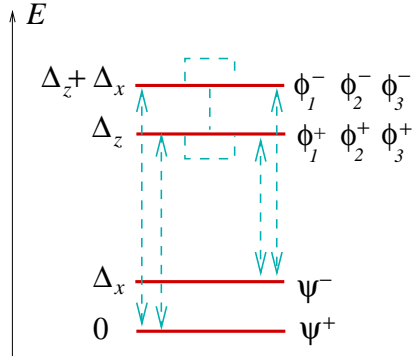}}
\caption{Allowed transitions induced by perturbation $V$ acting on
the eigenstates of $H$ are indicated by dashed lines. Direct
transitions between $\Psi^+$ and $\Psi^-$ levels are impossible.}
\end{figure}

We can use the expansion of Eq.~(\ref{self_energy}). The
perturbation $V$ is designed such that $V_{--}=\la
\Psi^+|V|\Psi^+\ra=\helse$, see Eq.~(\ref{HV}). The contribution of
the second-order term is proportional to the identity operator (see
Appendix~\ref{app:gadgets} for details of the calculation):
\[
V_{-+}G_{+}V_{+-} =
-(3/4)\omega^2
 \left[ \Delta_z^{-1} + (\Delta_z+\Delta_x)^{-1}\right]\, I \equiv \Omega\, I.
\]
We can regard it as a shift of energy. Therefore
\be\label{third_order} \Sigma_-(z) = \Omega\, I + \helse + V_{-+}
G_+ V_{++} G_+ V_{+-} + \mbox{[higher order terms]}. \ee The key
feature of the gadget is that the perturbation $V$ cannot cause a
direct transition from the ground-state $\Psi^+$ to the first
excited state $\Psi^-$ (or vice versa).  Any direct transition maps
$\Psi^+$ into the high-energy band $\phi^\pm$ spanned by six states
$\phi^+_j$ and $\phi^-_j$ having energy of order $\Delta_z\gg
\Delta_x$, see Figure~1. Thus any third-order process follows the
following scheme:
\[
\Psi^+\to \phi^\pm \to \phi^\pm \to \Psi^+.
\]
Since the energy splitting $\Delta_x$ of the $\phi^{\pm}$ band is
much smaller than its absolute energy $\Delta_z$, one can use an
approximation in which the two intermediate Green's functions $G_+$
in Eq.~(\ref{third_order})
 are proportional to the identity operator,
$G_+(z)=(zI-H_+)^{-1}\approx -I/\Delta_z$ for any $z=O(1)$.
Within this approximation one has
\[
\Sigma_-(z)  - \Omega\, I \approx \helse + \frac1{\Delta_z^2} V_{-+}
V_{++} V_{+-} = \helse +  \frac1{\Delta_z^2}\, \la
\Psi^+|V^3|\Psi^+\ra \approx \helse -\frac{3\omega^3}{\Delta_z^2}
(B_1 \otimes B_2 \otimes B_3 + B_1^\dag \otimes B_2^\dag \otimes
B_3^\dag),
\]
which approximates $\htar$ since $\omega^3=\Delta_z^2$. An accurate
calculation of $\Sigma_-(z)$, performed in
Appendix~\ref{app:gadgets}, shows that the error in the
approximation is of order $O(\delta)$. Contributions from
transitions involving the $\Psi^-$ level appear only in the
fourth-order term in Eq.~(\ref{self_energy}) according to the
following scenario: \be\label{fourth-order} \Psi^+\to \phi^\pm \to
\Psi^- \to \phi^\pm \to \Psi^+, \ee see Figure~1. In
Appendix~\ref{app:gadgets} we show that the fourth-order term is of
order $O(\delta)$. Therefore $\Sigma_-(z)=\Omega\, I + \htar +
O(\delta)$ for any $z=O(1)$, and thus the ground-state energy of
$\htar$ is $\delta$-close to the ground-state energy of
$\tilde{H}-\Omega\, I$.

One can applying this gadget in parallel to each term in the
Hamiltonian Eq.~(\ref{reduced_3_local}) and obtain the desired
reduction to a special $3$-local stoquastic Hamiltonian.

\subsection{Reduction from special $3$-local to $2$-local Hamiltonians}

To simplify the discussion let us consider a special $3$-local stoquastic Hamiltonian
with a single $3$-qubit interaction:
\[
\htar=\helse-6 B_1 \otimes B_2 \otimes B_3,
\]
where $B_j$ are non-negative operators proportional to $X_j$ and
$\helse$ is a $2$-local stoquastic Hamiltonian. The $3$-qubit
interaction can be treated using the original three-qubit gadget
in~\cite{KKR:hamsiam}. This original gadget coincides with the
gadget defined in Eq.~(\ref{HV}) if one chooses $\Delta_x=0$. In
this case the zero-energy subspace of $H$  is $\calL_-=\calH_{\rm
data}\otimes \calM_-$, where $\calM_-$ is spanned by the mediator
qubit states $|000\ra$ and $|111\ra$. Note that $\tilde{H}$ is now a
$2$-local stoquastic Hamiltonian.

We can choose $\Delta_z=\delta^{-3}$ and $\omega=\delta^{-2}$. The
analysis performed in~\cite{KKR:hamsiam} implies that the
ground-state energy of $\tilde{H}=H+V$, see Eq.~(\ref{HV}), is
$\delta$-close to the ground-state energy of an effective
Hamiltonian
\[
H_{\rm eff}=\Omega\, I + \helse\otimes I_m -6 B_1 B_2 B_3 \otimes
X_m,
\]
where $I_m$ and $X_m$ act on the two dimensional subspace of the
mediator qubits spanned by $|000\ra$ and $|111\ra$ (regarded as
logical $|0\ra$ and $|1\ra$ states). The energy shift is $\Omega\, I
=-\delta^{-1}(B_1^2 + B_2^2 + B_3^2)$. Since  $H_{\rm eff}$ is a
stoquastic Hamiltonian, the Perron-Frobenius theorem implies that
its ground-state $|\Psi_0\ra$ can be chosen as a non-negative
vector. Then a state
\[
|\Psi_0'\ra=|\Psi_0\ra + (I\otimes X_m)\,
|\Psi_0\ra
\]
is also a non-negative ground-state of $H_{eff}$.  In addition, we
have $(I\otimes X_m) |\Psi_0'\ra =|\Psi_0'\ra$. Therefore $H_{\rm
eff}-\Omega\, I$ has the same ground-state energy as $\htar$. This
proves that the ground-state energies of $\tilde{H}-\Omega\, I$ and
$\htar$ are $\delta$-close. To deal with multiple $3$-qubit terms in
Eq.~(\ref{xxx}) one applies this three-qubit gadget in parallel to
every individual $3$-qubit term.

{\it Remark:} In the original three-qubit gadget the operators $B_j$
are required to be positive semi-definite in order to guarantee that
the ground-state of $H_{\rm eff}$ belongs to the sector where
$X_{m}$ has eigenvalue $+1$.

\section{Stoquastic LH-MIN and Classical Post-Selected Computation}
\label{sec:POSTBPP}

The main goal of this section is to examine the complexity of
stoquastic LH-MIN in the special case when the Hamiltonian possesses
a polynomial spectral gap (i.e., the spectral gap scales as
$1/p(n)$, where $n$ is the number of qubits and $p$ is a fixed
polynomial). We shall prove that this problem can be placed in the
complexity class \POSTBPP{} --- a class of languages recognizable by
a probabilistic polynomial time classical circuits with a
post-selective readout of the answer. Speaking informally, any
problem in the class \POSTBPP{} can be solved by a classical
probabilistic circuit that outputs two random bits: $a$ (the answer
bit) and $b$ (the success flag). The answer bit $a$ contains the
correct answer of the problem provided that $b=1$ (if $b=0$ the
value of $a$ may be arbitrary). The success probability
$\prob{\left[ b=1\right]}$ must be positive for all input strings
(however it may be exponentially small). Here is a more formal
definition:

\begin{defi}[{\rm PostBPP}]
A promise problem $L=L_{\yes}\cup L_{\no}$ belongs to the class
\POSTBPP{} iff there exist a polynomial $p$, predicates $a(x,y)$ and
$b(x,y)$ from the class \P{} defined for any
 $y\in \Sigma^{p(|x|)}$, such that
\bea
x\in L &\implies& \prob{\left[ b(x,y)=1\right]}>0, \nn \\
x\in L_{\yes} &\implies & \prob{\left[ a(x,y)=1 \, |\, b(x,y)=1\right]}\ge 2/3, \nn \\
x\in L_{\no} &\implies & \prob{\left[ a(x,y)=1 \, |\,
b(x,y)=1\right]}\le 1/3. \nn \eea where $y\in \Sigma^{p(|x|)}$ is a
random uniformly distributed bit string, and $\prob{\left[a\, |\,
b\right]}$ is the conditional probability. \label{def:PostBPP}
\end{defi}

The quantum version of this class, PostBQP, was defined in Ref.
\cite{aaronson:postpp} and in that paper it was shown that
PP=PostBQP. The following lemma provides a characterization of
\POSTBPP{} in terms of the standard complexity classes.

\begin{lemma}
\label{lemma:path} $\MA \subseteq \NP^{\BPP} \, \subseteq
\POSTBPP=\BPPpath{} \subseteq \BPP^{\NP}\, \subseteq \Sigma^p_3$.
\end{lemma}
Here \BPPpath{} is a class of problems solvable in polynomial time
with a bounded error probability by a non-deterministic Turing
machine that chooses its computational path randomly from the
uniform distribution on a set of all possible paths,
see~\cite{Han:threshold}. The class \BPPpath{} is more powerful than
\BPP, since it offers the possibility to amplify the total
probability of successful computational paths by adding `idle'
computational branches to a non-deterministic algorithm. In Appendix
\ref{appb} we give a proof of the equality \POSTBPP=\BPPpath{}. All
other statements made in the previous lemma follow directly
from~\cite{Han:threshold}.

\begin{theorem}
\label{thm:POSTBPP} k-local stoquastic ${\rm LH-MIN}$ with the
promise that the spectral gap $\Delta=1/{\rm poly}(n)$ belongs to
\POSTBPP. \label{theo:post}
\end{theorem}

\begin{proof}
Let $H=\sum_S H_S\in \Omega(k,p_1,p_2)$ be $k$-local stoquastic
Hamiltonian on $n$ qubits, see Definitions~\ref{def:LHP1} and
\ref{def:LHP2}. The first step is to transform $H$ into a
doubly-substochastic\footnote{ By definition, a non-negative matrix
is doubly-substochastic iff the sum of the
elements in every row and every column is smaller or equal to $1$,
see~\cite{book:bhatia}} matrix $G$. This is achieved by choosing
\be
\label{eq:q}
G=\frac12\, (I- H/q(n)), \quad q(n)=2 \max(1, 2^k {n\choose k}
p_1(n)).
\ee
The choice of $q(n)$ in Eq.~(\ref{eq:q}) takes into account that
$H$ contains at most ${n\choose k}$ local terms $H_S$ and each
local term $H_S$ has at most $2^k$ non-zero matrix elements in any
row (column).  This choice of $q(n)$
also guarantees that all eigenvalues of $G$
are between $0$ and $1$, while the matrix elements $G_{x,y}=\la
x|G|y\ra$ obey the inequalities \be \label{bounds on b} G_{x,y}\ge 0
\quad \mbox{and} \quad \frac14 \le \sum_{z\in \Sigma^n} G_{x,z}\le 1
\quad \mbox{for all} \quad x,y\in \Sigma^n. \ee
Obviously,  $q(n)$
is a fixed polynomial.
  Let $\mu(G)$ be the largest eigenvalue of $G$.
The correct decision for LH-MIN with the Hamiltonian $H$ can be made
if we can evaluate $\mu(G)$ with polynomial precision:
\begin{eqnarray}
\lambda(H)\le 0 &\slno& \mu(G)\ge \mu_+ = \frac12, \nn \\
\lambda(H)\ge 1/p_2(n) &\slno & \mu(G) \le \mu_-=\frac12\left( 1-
\frac1{q(n)p_2(n)}\right). \nn
\end{eqnarray}

We shall present a polynomial-time probabilistic algorithm that
evaluates $\mu(G)$ with a precision $1/{\rm poly}(n)$ using a
post-selective readout of the answer.

Define a matrix $B$ which is diagonal in the standard basis such
that \be \label{B_x} B_x\equiv \la x|B|x\ra = \sum_{y\in \Sigma^n}
G_{x,y}. \ee We can transform $G$ into a doubly-stochastic matrix
$F$ as follows:
\[
F=G\otimes I + (I-B)\otimes X = \left( \ba{cc} G & I-B \\ I-B
& G \\ \ea \right).
\]
The matrix $F$ acts on $n$ original qubits and one extra ancillary qubit. The states $|0\ra$ and $|1\ra$ of the
ancillary qubit label the four blocks  in the matrix representation of $F$ given above.
The purpose of the ancillary qubit is to enlarge the space of states of the random walk such that for every under-normalized row
of $G$ the walker can "leak" to one of the ancillary states (those in which the ancillary qubit is $|1\ra$)
thus making the corresponding row of $F$ normalized.
Therefore $F$ specifies a random walk on a
space $\Sigma^{n+1}$. The fact that $H$ is a $k$-local Hamiltonian
implies that $F$ is a sparse matrix --- it has at most ${n\choose k}
2^k+1$ non-zero elements in each column (row). Moreover, for any
fixed column (row) positions of the  non-zero matrix elements and
their values can be computed in ${\rm poly}(n)$ time. This means
that the random walk defined by $F$ can be efficiently simulated on
a \BPP{} machine, provided that the number of steps is at most ${\rm
poly}(n)$.

Our algorithm requires the simulation of $w$ independent random
walks $(X^{(i)}_t)_{t=0,\dots,L,i=1,\dots,w}$ whose transition
probabilities are given by $F$. Here $0\leq t \leq L$ is the
(discrete) time parameter, $i$ is the index of the random walk and
$L$, $w$ will be specified later. Let us start each random walk
$X^{(i)}_t$ from a point $X^{(i)}_0=(x^{(i)}_0,0)\in \Sigma^{n+1}$,
such that the ancillary bit (the last one) is set to $0$, and the
$n$ bits constituting the original system are initialized by a
random string $x^{(i)}_0\in \Sigma^n$ drawn from the uniform
distribution with independent choices of $x^{(i)}_0$ for different
$i$. Suppose that after $t$ steps the $i$th random walk arrives at a
point $X^{(i)}_t=(x^{(i)}_t,b^{(i)}_t)$ ($0\leq t\leq L$, $1\leq
i\leq w$). Let us postselect only those samples where the ancillary
bits remain in the state $0$ for the whole duration of each of the
$w$ walks. In terms of the formal definition of \POSTBPP{} we have
to define the success flag bit as $b=\neg(\vee_{i=1}^w\vee _{t=0}^L
b^{(i)}_t)$. The probability for the ancillary bit to stay in $0$ is
\[
\prob{\left[ b=1\right]} =\left(\frac1{2^n} \sum_{x_0,x_L\in \Sigma^n} \la
x_0|G^L|x_L\ra \right)^w \ge \frac1{4^{wL}}>0,
\]
where we have used the inequality Eq.~(\ref{bounds on b}).

Conditioned on $b=1$, the random variables $(x^{(i)}_L)_{i=1}^w$ are independent samples from the probability distribution $P_L(\cdot)$ given by
\[
P_L(y)=\frac{\sum_{x\in \Sigma^n} \la x|G^L|y\ra }{ \sum_{x,y\in
\Sigma^n} \la x|G^L|y\ra},\; y\in\Sigma^n.
\]
Consider a quantity \be\label{eq:estimateformu}\mu_{\rm
est}(G)\equiv \frac{\sum_{i=1}^{w}B_{x^{(i)}_L}}{w} =
\frac{\sum_{i=1}^{w}\left(\sum_{x\in \Sigma^n}\la
x|G|x^{(i)}_L\ra\right)}{w}\ee
Given the samples $x^{(i)}_L$, the quantity $\mu_{\rm
est}(G)$ can be efficiently computed since $G$ is a sparse matrix.

The expectation value of  $\mu_{\rm est}(G)$
taken over the $w$ independent samples of $x^{(i)}_L$)
is equal to
$$\ex{(\mu_{\rm est}(G))} = \frac{\sum_{x,y\in \Sigma^n} \la x|G^{L+1}|y\ra}{\sum_{x,y\in
\Sigma^n} \la x|G^L|y\ra}.$$ Since $1/4\leq B_x\leq 1$ for all
$x\in\Sigma^n$, Azuma's inequality implies that
\be\label{eq:goodapprox}\forall \delta>0,\;\;\prob{\left(\left.
\left|\,\mu_{\rm est}(G)-\ex{(\mu_{\rm
est}(G))}\right|>\delta\,\right|\, b=1\right)}\leq
2e^{-\frac{\delta^2w}{2}}.\ee We now {\em claim} that for $L$ chosen
sufficiently large,
$$\ex{(\mu_{\rm est}(G))}=\frac{\sum_{x,y\in \Sigma^n} \la x|G^{L+1}|y\ra}{\sum_{x,y\in
\Sigma^n} \la x|G^L|y\ra}$$ is
close to the largest eigenvalue $\mu(G)$ of $G$. More precisely, we will show that
\begin{lemma}
\label{lemma:gap} Let $\mu_0 \geq \mu_1$ be the largest eigenvalue
and second largest eigenvalue of $G$. Suppose that
$\log{(\mu(G))}-\log{(\mu_1)} \ge \frac{1}{r(n)}$. If one chooses
$L=\frac{5 nr(n)}{2}$ then
\[
|\mu_0-\ex{(\mu_{\rm est}(G))}| = O(2^{-n}).
\]
\end{lemma}
This is the only step where the spectral gap assumption is used. Let
us postpone the proof of the lemma until the end of the section. We
choose $L=\frac{5 n r(n)}{2}$ (clearly, the spectral gaps of $H$ and
$G$ are related by a polynomial factor, so that $r(n)$ is a fixed
polynomial). Then by Eq.~(\ref{eq:goodapprox}) \be\nonumber \forall
\delta>0,\;\;\prob{\left(\left.|\mu_{\rm est}(G)-\mu(G)|>\delta +
O(2^{-n})\,\right|\, b=1\right)}\leq 2e^{-\frac{\delta^2w}{2}}.\ee
For some constant $c>0$, taking $w=2n^{2c}\ln(6)$ ensures that
$|\mu_{\rm est}(G)-\mu(G)|=O(n^{-c})$ with probability at least
$2/3$. Since the coefficients $B_y$ can be computed efficiently for
any bit-string $y$ and vary within a constant range we can evaluate
$\mu_{\rm est}(G)$ (as in Eq. (\ref{eq:estimateformu})) with a
precision $1/{\rm poly}(n)$ using $w={\rm poly}(n)$ random walks of
length $L={\rm poly(n)}$. The complexity of simulating the random
walks is polynomial in $L$, $w$, and $n$; it follows that we can
solve our decision problem in \POSTBPP{}. \end{proof}

\noindent
{\bf Proof of Lemma~\ref{lemma:gap}:}\\
Define an operator
\[
\hat{\Delta}_\ell=\frac1{\mu_0^\ell}\left(G^\ell - \mu_0^\ell \, |\Psi_0\ra\la
\Psi_0|\right).
\]
After simple algebra one gets
\[\ex{(\mu_{\rm est}(G))}=
\mu_0\left(\frac{1 + \epsilon^2 \sum_{x,y\in\Sigma^n}\la
x|\hat{\Delta}_{L+1}|y\ra}{1 + \epsilon^2 \sum_{x,y\in\Sigma^n}\la
x|\hat{\Delta}_{L}|y\ra}\right), \quad \epsilon \equiv \frac1{\sum_x
\la x|\Psi_0\ra}.\]
Let $\mu_0\geq \mu_1 \geq \ldots,\mu_{2^n-1}$ be
the eigenvalues of $G$. Note that $G$ is chosen such that $\mu_j\ge
0$. Therefore
\[
|| \hat{\Delta}_L || =\left(\frac{\mu_1}{\mu_0}\right)^L \le
2^{-\frac{L}{r(n)}}.
\]
Let us choose $L=\frac{5 n r(n)}{2}$. Then $|| \hat{\Delta}_L ||\le
2^{-5n/2}$ and therefore
\[\left|
\sum_{x,z} \la x|\hat{\Delta}_L|z\ra\right|  \le 2^{n}
\left|\left(\frac{\sum_{x}\la x|}{2^{n/2}}\right)
\hat{\Delta}_L\left(\frac{\sum_{z}|z\ra}{2^{n/2}}\right) \right| \le
2^{n}||\Delta_L||\leq 2^{-3n/2}.\] Clearly, the same inequalities
hold with $L+1$ replacing $L$. On the other hand, $\epsilon \leq 1$
since
$$\sum_{x\in\Sigma^n}\la x|\Psi_0\ra \geq \sqrt{\sum_{x\in\Sigma^n}(\la x|\Psi_0\ra)^2}=1.$$
It follows that $|\mu_0-\ex{(\mu_{\rm est}(G))}| \leq O(\mu_0
2^{-3n/2})=O(2^{-n})$ as $\mu_0\leq \max_x B_x \leq 1$ under our
assumptions.\qed

Our result has a simple implication for adiabatic quantum
computation using stoquastic Hamiltonians. It is known that the
power of efficient adiabatic quantum computation with general
2-local Hamiltonians is equal to that of polynomial-time quantum
circuits \cite{ADLLKR:adia}. All Hamiltonians on the adiabatic path
are required to have a polynomial gap in order for the adiabatic
theorem to apply. Now let us restrict ourselves to stoquastic
Hamiltonians with a polynomial gap. By the MA-hardness construction
and analogous to the arguments in \cite{ADLLKR:adia}, one can argue
that any polynomial-time probabilistic computation can be simulated
by an efficient adiabatic path using stoquastic Hamiltonians only.
It is a more interesting but open question whether every efficient
adiabatic path using stoquastic Hamiltonians can be simulated by a
polynomial-time probabilistic machine. The proof of Theorem
\ref{theo:post} shows that post-selected classical computation
allows one to efficiently sample from the ground-state distribution
of a stoquastic Hamiltonian. Note that this may be potentially
stronger than merely estimating the lowest-lying eigenvalue. In the
proof we use the ability to sample from the ground-state to estimate
the lowest-lying eigenvalue. A adiabatic path with stoquastic
Hamiltonians, each of which has a $1/{\rm poly}(n)$ gap, can thus be
simulated by post-selected classical computation and the decision
problem that can be solved by these means is contained in ${\rm
PostBPP}$.

\section{Acknowledgements}

We acknowledge support by the NSA and the ARDA through ARO contract
number W911NF-04-C-0098.

\appendix

\section{The Approximate Counting Problem and Hash Functions}
\label{app:a}

For the sake of completeness we explain how to choose the parameters
of the hash functions in the proof of Theorem \ref{thm:AM}, see the
original paper~\cite{GS:coin} for more details. Define
\[
b=\lceil \log{\lrg} \rceil +3.
\]
Without loss of generality  $b\le k$ (otherwise Arthur has to verify
that $\Omega$ contains a finite fraction of $k$-bit strings, which
can be done by the standard Monte-Carlo method without compression).
Let $h_1,\ldots,h_k$ be $k\times b$ binary matrices chosen uniformly at random.
Each matrix $h_j$ defines a linear hash function
$h_j\, : \, \Sigma^k \to \Sigma^b$. Denote
\[
h(\Omega) = \bigcup_{j=1}^k h_j(\Omega) \subseteq \Sigma^b.
\]
We need the following technical lemma from~\cite{GS:coin} (a proof
is given at the end of this appendix).
\begin{lemma}
\label{lemma:prob} For any set $\Omega\subseteq \Sigma^k$ and for
any $b\le k$ such that $|\Omega|\le 2^{b-2}$ one has
\[
\prob{\left[|h(\Omega)|\ge \frac{|\Omega|}k \right]}\ge
1-\frac1{2^k}.
\]
\end{lemma}
Neglecting the exponentially small error probability $2^{-k}$ one
gets \bea |\Omega|\ge \lrg &\implies &
|h(\Omega)|\ge \frac{\lrg}k \ge  \left(\frac1{8k}\right) 2^b, \nn \\
|\Omega|\le \sml &\implies & |h(\Omega)|\le k\cdot \sml \le
k2^{-n}\, \lrg \le \left( \frac{k}{2^{n+2}}\right) 2^b \nn \eea For
the second line we have used the trivial bound $|h(\Omega)|\le
k|\Omega|$ and Eq.~(\ref{ls1}). If $n$ is sufficiently large,
$h(\Omega)$ contains a polynomially large fraction of $b$-bit strings
for positive instances and an exponentially small fraction for negative
instances. Arthur can distinguish the two case by the Monte-Carlo
method using Merlin's advice to verify membership in $h(\Omega)$.
This completes the proof of Theorem~\ref{thm:AM}. \qed

\noindent
{\bf Proof of Lemma~\ref{lemma:prob}:}\\
Let us say that a function  $h_j$ is {\it invertible } at the point
$x\in \Omega$ if $h_j(x)\ne h_j(y)$ for all $y\in \Omega\backslash
\{x\}$. Define a set
\[
\Omega_j =\{ x\in \Omega\, :\, h_j \; \mbox{is invertible at} \;
x\}.
\]
Clearly,
\[
|h(\Omega)|\ge |h_j(\Omega)|\ge |\Omega_j| \quad \mbox{for any}
\quad j=1,\ldots,k.
\]
Thus \be \label{aux1} \prob{\left[|h(\Omega)|\ge \frac{|\Omega|}k
\right]}\ge \prob{\left[ \bigcup_{j=1}^k \Omega_j =\Omega\right]}.
\ee Since the probability of collisions for $h_j$ is $2^{-b}$, we
have
\[
\prob{\left[h_j \; \mbox{is not invertible at} \; x\right]} \le
\frac{|\Omega|}{2^b}.
\]
Therefore \be \label{aux2} \prob{\left[ \bigcup_{j=1}^k \Omega_j \ne
\Omega\right] } = \prob{\left[ \exists\,  x\in \Omega \, : \,
\forall j \; h_j \; \mbox{is not invertible at} \; x\right]} \le
|\Omega| \left(\frac{|\Omega|}{2^b}\right)^k. \ee Combining
Eqs.~(\ref{aux1}) and (\ref{aux2}) and taking into account the conditions
on $b$, $k$, and $|\Omega|$ finishes the proof. \qed

\section{${\rm PostBPP}={\rm BPP}_{\rm path}$}
\label{appb}

The class \BPPpath{} is defined most conveniently in terms of
non-deterministic Turing machines. Let $M$ be a non-deterministic
Turing machine (TM). We shall assume that at each step $M$ chooses
one of two computational paths. Given an input string $x\in
\Sigma^*$, a polynomial-time non-deterministic TM makes at most
$q(|x|)$ steps before it stops, where $q$ is a fixed polynomial.
Whenever $M$ stops, it outputs an answer bit $a=1$ (accept), or
$a=0$ (reject).

Let $\paths{x}$ and $\acc{x}\subseteq \paths{x}$ be a set of all
computational paths and a set of accepting paths for a machine $M$
running on input string $x$. By definition, $|\paths{x}|\le
2^{q(|x|)}$. One can visualize $\paths{x}$ as a subtree of a binary
branching tree of a height $q(|x|)$. Some paths make it all the way
from the root to a leaf of the tree and some paths end before
making $q(|x|)$ steps. Let us introduce a branching variable $y\in
\Sigma^{q(|x|)}$, such that a bit $y_j$ specifies what path $M$
chooses at step $j$ (if a computational path ends before making
$q(|x|)$ steps, the remaining bits of $y$ can be ignored). For any
$x\in \Sigma^*$ and $y\in \Sigma^{q(|x|)}$ let $l(x,y)$ be the
number of steps that $M$ does on input $x$ before it stops and
$a(x,y)$ be the value of the answer bit. By definition, $1\le
l(x,y)\le q(|x|)$ for any $x,y$ and \be \label{paths}
|\paths{x}|=\frac1{2^{q(|x|)}} \sum_{y\in \Sigma^{q(|x|)}}
2^{l(x,y)}, \quad |\acc{x}|=\frac1{2^{q(|x|)}} \sum_{y\, : \,
a(x,y)=1} 2^{l(x,y)}. \ee Now we can define the class \BPPpath{}
more formally.
\begin{defi}
\label{def:path} A promise problem $L=L_{\yes}\cup L_{\no}$ belongs
to the class \BPPpath{} iff there exist a non-deterministic
polynomial-time Turing machine $M$ such that \bea
x\in L_{\yes} &\implies & |\acc{x}|\ge \frac23\, |\paths{x}| \nn \\
x\in L_{\no} &\implies & |\acc{x}|\le \frac13\, |\paths{x}| \nn \eea
\end{defi}

Let us first prove $\BPPpath\subseteq \POSTBPP$. Indeed, consider
a non-deterministic polynomial-time Turing machine $M$ as above. Let
$C$ be  a classical circuit (more strictly, a uniform family of
circuits) that takes as input a pair $(x,y)$ with $y\in
\Sigma^{q(|x|)}$, and simulates $M$ for $q(|x|)$ steps according to
the computational path $y$. The circuit $C$ outputs the answer bit
$a(x,y)$ and the number of steps $l(x,y)$ in the path $y$. The idea
is that we can simulate $M$ by choosing $y$ randomly from the
uniform distribution and use post-selection to balance the resulting
distribution on $\paths{x}$. Indeed, define a random success flag
bit $b$, such that we have a probability distribution of $b$ conditioned on
$x$ and $y$
\[
\prob{\left[b=1\, |\, x,y\right]}=\frac1{2^{q(|x|)-l(x,y)}}.
\]
Since the circuit $C$ outputs $l(x,y)$, one can easily generate a
bit with the desired distribution using a polynomial number of
ancillary random bits.
 Making use of the formulas in Eq.~(\ref{paths}) one can easily get
\[
\prob{\left[ a=1\, |\, b=1\right]} = \frac{\prob{\left[
a=1,b=1\right]}}{\prob{\left[ b=1\right]}} = \frac{2^{-2\,
q(|x|)}\sum_{y\, : \, a(x,y)=1} 2^{l(x,y)}}{ 2^{-2\, q(|x|)}\sum_{y}
2^{l(x,y)}} = \frac{ |\acc{x}|}{|\paths{x}|}.
\]
Comparing it with Def.~\ref{def:PostBPP}, we conclude that a
language recognized by $M$ belongs to \POSTBPP.

Now let us prove $\POSTBPP\subseteq \BPPpath$.  Indeed, let
$L=L_{\yes}\cup L_{\no}$ be a language from $\POSTBPP$. One can use
the standard majority voting procedure to reduce the error
probability from $1/3$ to $1/4$, i.e., we can assume that the
predicates $a(x,y)$ and $b(x,y)$ from Def.~\ref{def:PostBPP} satisfy
\bea
x\in L &\implies& \prob{\left[ b(x,y)=1\right]}>0, \nn \\
x\in L_{\yes} &\implies & \prob{\left[ a(x,y)=1 \, |\, b(x,y)=1\right]}\ge 3/4, \nn \\
x\in L_{\no} &\implies & \prob{\left[ a(x,y)=1 \, |\,
b(x,y)=1\right]}\le 1/4. \nn \eea Here $y\in \Sigma^{p(|x|)}$ is a
uniformly random bitstring and $p$ is a polynomial. The
inequality $\prob{\left[b=1\right]}>0$ implies that there exists at
least one $y\in \Sigma^{p(|x|)}$ such that $b(x,y)=1$. Therefore we
can bound the probability of successful computation from below as
\[
\prob{\left[b=1\right]}\ge \frac1{2^{p(|x|)}}.
\]
Construct a non-deterministic Turing machine $M$ that takes $x$ as
input
and does the following:\\
(1) Perform $p(|x|)$ branchings to initialize a string $y\in \Sigma^{p(|x|)}$,\\
(2) Compute predicates $a=a(x,y)$ and $b=b(x,y)$, \\
(3) If $b=0$, output $a$,\\
(4) If $b=1$, perform $p(|x|)+4$ idle branchings and output $a$.\\
Let us verify that $M$ recognizes the language $L$ in the sense of
Def.~\ref{def:path}. Indeed, one can easily check that
\[
|\paths{x}|=2^{p(|x|)}\left[ \prob{[b=0]} + 2^{p(|x|)+4}\,
\prob{[b=1]}\right]
\]
and
\[
|\acc{x}|=2^{p(|x|)}\left[ \prob{[a=1,b=0]} + 2^{p(|x|)+4}\,
\prob{[a=1,b=1]}\right].
\]
Consider first the case $x\in L_{\yes}$. Then
\[
\frac{|\acc{x}|}{|\paths{x}|}\ge
\frac{\prob{[a=1,b=1]}}{2^{-p(|x|)-4} + \prob{[b=1]}}\ge
\frac{\prob{[a=1,b=1]}}{\prob{[b=1]}(1+2^{-4})} \ge \frac3{4(1+
2^{-4})} > \frac23.
\]
Here we have used the fact that $\prob{[b=1]}\ge 2^{-p(|x|)}$.
Consider now the case $x\in L_{\no}$. Then
\[
\frac{|\acc{x}|}{|\paths{x}|}\le \frac{2^{-p(|x|)-4} +
\prob{[a=1,b=1]}}{\prob{[b=1]}} \le \frac14 + 2^{-4} <\frac13.
\]
Thus $M$ indeed recognizes $L$. \qed

\section{The three-qubit gadget}
\label{app:gadgets}

In this appendix we will explicitly calculate the self-energy
operator $\Sigma_-(z)$ for the perturbed Hamiltonian in
Eq.~(\ref{HV}) up to third order in the perturbative series of
Eq.~(\ref{self_energy}). We shall also evaluate the norm of the
fourth-order term.
It follows directly from Eq.~(\ref{HV}) that
\[
V_{--}=\la\Psi^+|V|\Psi^+\ra = H_{\rm else}.
\]
A straightforward calculation yields
\[
V_{+-}=-\frac{\omega}2 \sum_{j=1}^3 \sum_{\alpha=\pm 1}
(B_j + \alpha B_j^\dag)\otimes |\phi^\alpha_j\ra,
\]
where $|\phi^\alpha_j\ra=\sigma^x_j |\Psi^\alpha\ra$, see
Eq.~(\ref{eigen_basis}). Now we can compute the second-order term
for the self-energy operator: \be\label{2nd} \Sigma_-^{(2)}(z)=
V_{-+}G_+V_{+-} =\left(-\frac{\omega}2\right)^2 \sum_{j=1}^3
\sum_{\alpha=\pm 1} \frac{ (B_j^\dag + \alpha B_j)(B_j + \alpha
B_j^\dag)}{z-\Delta_\alpha}, \ee Here we denote $\Delta_+=\Delta_z$
and $\Delta_-=\Delta_z+\Delta_x$. Substituting $z=O(1)$ and taking
into account that $B_j^2=0$, $B_j B_j^\dag + B_j^\dag B_j =I$, we
come to \be\label{shift} \Sigma_-^{(2)}(z)= \Omega\, I +
O(\delta^4), \quad \Omega = -(3/4)\omega^2
 \left[ \Delta_z^{-1} + (\Delta_z+\Delta_x)^{-1}\right].
\ee To compute the third-order term we need to know $V_{++}$. It is
enough to find the matrix elements of $V$ between the $\phi^{\pm}$
states (since transitions between $\Psi^-$ and $\phi^{\pm}$ do not
appear in the third order). A straightforward calculation yields \be
\la \phi^\alpha_j |V|\phi^\beta_l\ra = H_{\rm else}
\delta_{j,l}\delta_{\alpha,\beta} -\frac{\omega}2 \sum_{k=1}^3
\epsilon(j,k,l) \left[ \alpha B_k + \beta B_k^\dag\right], \quad
\mbox{where} \quad \epsilon(j,k,l)=\left\{ \ba{rcl}
1 &\mbox{if}& j\ne k\ne l \\
0 &&\mbox{otherwise} \\
\ea
\right.
\ee
Therefore
\[
\Sigma^{(3)}_-(z)=
V_{-+}G_+V_{++}G_+V_{+-}=
\left(-\frac{\omega}2\right)^3 \sum_{j,k,l=1}^3 \sum_{\alpha,\beta=\pm 1}
\frac{(B_j^\dag + \alpha B_j)(\alpha B_k + \beta B_k^\dag)(B_l + \beta B_l^\dag)\epsilon(j,k,l)}
{(z-\Delta_\alpha)(z-\Delta_\beta)} + O(\delta^4).
\]
Taking into account Eq.~(\ref{scaling}) one easily gets (for any
$z=O(1)$) \be \Sigma^{(3)}_-(z)=-3(B_1 \otimes B_2 \otimes B_3 +
B_1^\dag \otimes B_2^\dag \otimes B_3^\dag) + O(\delta). \ee
Although we have not calculated the fourth-order correction
\[
\Sigma_-^{(4)}=
V_{-+}G_+V_{++}G_+V_{++}G_+ V_{+-}
\]
exactly, we have to get an upper bound on its norm. The fourth-order
processes may involve the low-lying level $\Psi^-$, see
Eq.~(\ref{fourth-order}), and potentially these processes can give a
non-negligible contribution to $\Sigma_-$ as $\Sigma_-^{(3)}$.
Keeping in mind Eq.~(\ref{fourth-order}) one can easily get (for any
$z=O(1)$)
\[
||\Sigma_-^{(4)}(z)|| =||V_{-+}G_+ V_{++} G_+ V_{++} G_+ V_{+-}||
=O\left(\frac{\omega^4}{\Delta_z
\Delta_x\Delta_z}\right)=O(\delta^{-16+12+5})=O(\delta).
\]
As for the higher-order corrections to $\Sigma^-$ (from the fifth-
order onwards) their contribution contains an additional factor
$\omega/\Delta_z$, or $\omega/\Delta_x$ which is at most $\delta$.
Therefore we arrive at
\[
\Sigma_-(z)=\Omega\, I + H_{\rm else} -3(B_1 \otimes B_2 \otimes B_3
+ B_1^\dag \otimes B_2^\dag \otimes B_3^\dag) + O(\delta)
\]
for any $z=O(1)$. Here $\Omega$ is the energy shift given by Eq.~(\ref{shift}).

\bibliographystyle{abbrv}

\end{document}